\documentclass[12pt]{article}
\usepackage{etex}
\usepackage{epsfig}
\usepackage{epsf}
\usepackage{latexsym}
\usepackage{amsfonts,amssymb,amsmath} 
\usepackage{mathtools,cancel}
\usepackage{color}
\usepackage[a4paper,vmargin=3cm,hmargin=2.1cm]{geometry}
\usepackage{tikz}
\usepackage[all]{xy}
\epsfverbosetrue
\newcommand{\goth}[1]{\mathfrak{#1}}

\newcommand{\aaa}{\mathcal{A}}
\newcommand{\aaab}{\mathcal{\bar A}}
\newcommand{\bbb}{\mathcal{B}}
\newcommand{\ccc}{\mathcal{C}}
\newcommand{\ssss}{\goth{s}}

\newcommand{\sss}{\mathcal{S}}
\newcommand{\ssst}{\mathcal{\tilde S}}
\newcommand{\rrr}{\mathcal{R}}
\newcommand{\zzz}{\goth{z}}

\newcommand{\de}{\hbox{\rm{d}}}

\newcommand{\pa}{\partial}

\newcommand{\lb}{\left[}
\newcommand{\rb}{\right]}
\newcommand{\lp}{\left(}
\newcommand{\rp}{\right)}
\newcommand{\la}{\left\{}
\newcommand{\ra}{\right\}}

\newcommand{\dpp}{\vcentcolon}
\newcommand{\bb}{\begin{eqnarray}}
\newcommand{\ee}{\end{eqnarray}}
\newcommand{\eee}{\nonumber\end{eqnarray}}
\newcommand{\qq}{\quad}

\newcommand{\nc}{\newcommand}
 \nc{\alf}{\alpha} 
\nc{\ze}{\zeta}
\nc{\tht}{\theta} \nc{\T}{\Theta} \nc{\be}{\beta}  \nc{\eps}{\epsilon} 
\nc{\ga}{\gamma}  \nc{\De}{\Delta} 
 \nc{\G}{\Gamma}  \nc{\vphi}{\varphi}
 \nc{\si}{\sigma}  \nc{\ka}{\kappa}   \nc{\Si}{\Sigma} 
\nc{\om}{\omega}  \nc{\chic}{\widehat{\chi}}

\nc{\qqq}{\quad\quad}               

 \nc{\Om}{\Omega}
\nc{\nf}{\infty}   \nc{\dl}{\mathop{\smash{\cal L}}}  \nc{\black}{\rule{3mm}{3mm}}
\nc{\ol}{\overline}        \nc{\und}{\underline} 
\nc{\beq}{\begin{equation}}  \nc{\eeq}{\end{equation}}  \nc{\pt}{\partial}  
   \nc{\dst}{\displaystyle}  \nc{\na}{\nabla} 
\nc{\nnb}{\nonumber}    \nc{\bs}{\backslash}        \nc{\mb}{\mathbb}   
\nc{\sn}{{\rm sn}\,} \nc{\cn}{{\rm cn}\,}     \nc{\dn}{{\rm dn}\,} \nc{\nin}{\noindent}
\nc{\ti}{\tilde}   \nc{\wti}{\widetilde}   \nc{\h}{\hat}  \nc{\wh}{\widehat}
\nc{\tpsi}{\wti{\psi}}   \nc{\tphi}{\wti{\phi}}  \nc{\tH}{\wti{H}} 
\nc{\arc}{{\rm arc}\,}

\newcounter{muni}
\newenvironment{remunerate}{\begin{list}{{\rm \arabic{muni}.}}
{\usecounter{muni}
\setlength{\leftmargin}{0pt}\setlength{\itemindent}{38pt}}}{\end{list}}

\nc{\brm}{\begin{remunerate}}   \nc{\erm}{\end{remunerate}}

\nc{\stg}{\mathop{\smash{*}}}
\nc{\st}{\mathop{\smash{\delta}}}
\nc{\barr}{\begin{array}}   \nc{\earr}{\end{array}}   \nc{\dg}{\dagger}
\nc{\mtvb}{\mathversion{bold}}   \nc{\mtvn}{\mathversion{normal}}

\begin{document}

\thispagestyle{empty}

\begin{center}
${}$
\vspace{3cm}

{\Large\textbf{Axial Bianchi IX meets Gaia data}} \\

\vspace{2cm}

{\large

Thomas Sch\"ucker\footnote{
Aix Marseille Univ, Universit\'e de Toulon, CNRS, CPT,  Marseille, France\\\indent\qq
thomas.schucker@gmail.com }
}

\vspace{3cm}

{\large\textbf{Abstract}}
\end{center}
The aim of the present work is two-fold: 

{\it(i)} Compute the drift of comoving extragalactic sources in axial Bianchi IX universes, in particular in those satisfying Einstein's equations with positive cosmological constant and comoving dust. 

{\it(ii)} Linearize this drift (and for comparison the previously calculated Lema{\^i}tre-Hubble diagram) simultaneously in small anisotropy and in small positive curvature.

 We find that the drift in this linear approximation coincides with the drift in axial Bianchi I universes. This is not true for Lema{\^i}tre-Hubble diagrams.

\vspace{1cm}

\hfill{\em To the memory of Andr\'e Tilquin}

\vspace{1cm}

\vspace{3cm}


\vskip 1truecm
\vfil\eject

\section{Introduction}

A simple theory we use to try and understand our universe is built on two pillars, general relativity and the cosmological principle. Today more and more precise observations  challenge this theory. One option for gaining flexibility is by breaking the cosmological principle, which postulates maximal symmetry in 3-spaces of simultaneity. Since the simple theory works quite well we feel encouraged to explore {\it minimal} breakings of these 6-dimensional isometry groups. The necessary mathematics have been worked out by Luigi Bianchi  \cite{Bianchi} (1898) and his student Guido Fubini \cite{Fubini} (1903) well before the discovery of general relativity. They teach us that there are only three minimal symmetry breakings of the cosmological principle: take the Friedman universes of zero, negative or positive spatial curvature and deform them (differentiably) while keeping  4-dimensional isometry sub-groups unbroken. These are the axial Bianchi I, V and IX universes.

We say `Friedman universes' because we subsequently also want their metrics to satisfy Einstein's equations. The Lema{\^i}tre-Hubble diagram of supernovae and the drift of quasars concern our near past with redshift below 6. In this time interval the energy-momentum content of the simple theory mentioned above is dominated by comoving dust making the theory even simpler. At this stage we must acknowledge Farnsworth's astonishing theorem \cite{farn} from 1967: all axial Bianchi V universes are incompatible with Einstein's equations and comoving dust (unless these universes are already maximally symmetric). Valent gives an alternative proof of 
Farnsworth's theorem in the long version of \cite{vts}.

Therefore we are left with axial Bianchi I and IX universes.

Already in 1933 (tri-axial) Bianchi I universes were  analyzed by George Lema\^itre \cite{lemaitre} following a suggestion by Albert Einstein. Probably both were unaware of Bianchi's work. Today we know the exact solutions of Einstein's equations with comoving dust  in all Bianchi I universes \cite{stv}. We know the Lema{\^i}tre-Hubble diagram \cite{stv,fpu} and the drift of comoving extragalactic sources \cite{Mar,drift1}.

Axial Bianchi IX universes are minimal deformations of Friedman universes with positive spatial curvature and are motivated by a weak signal of small positive curvature in the Cosmic Microwave Background data \cite{uke,wh,div}. 

We completely agree with Will Handley's point of view \cite{wh}:
``Despite the long history of cosmological models which include spatial curvature, in the modern era there is a strong research community bias toward a flat universe. [...] Curved models are also likely under-represented in the literature due in large part to the increased theoretical and numerical computational cost associated with curved cosmologies.'' 

A similar feeling motivated us to compute the Lema{\^i}tre-Hubble diagram in axial Bianchi IX universes \cite{vts} and also motivates the present work.

We are also motivated by recent observations: The third Gaia data release (2022) \cite{gaia} contains drift measurements of over a million quasars with redshift between 0.5 and 3. Their fit to axial Bianchi I universes shows a significant discrepancy \cite{makarov} with Cea's fit  \cite{cea1} from 2014. He used the Cosmic Microwave Background data from the Wilkinson Microwave Anisotropy Probe and the 2013 Planck data. This discrepancy
raises the question whether axial Bianchi IX universes with their two additional parameters can modify the theoretical drift. We will show that they cannot.

In the limit of small curvature, curved Friedman universes tend {\it locally} to flat Friedman universes.
Similarly axial Bianchi IX universes   approach  axial Bianchi I universes.  After calculating the drift in axial Bianchi IX universes, we propose to take the double limit of small anisotropy and small curvature for Lema{\^i}tre-Hubble diagrams and drifts.

The outline of the paper is as follows:

Section \ref{kin} is purely kinematical and uses the solutions of the geodesic equations in axial Bianchi IX universes that we have computed in \cite{vts} using Noether's theorem with the four Killing vectors. Without the use of Einstein's field equations we compute the drift of a comoving extra-galactic source as seen by a comoving observer. These results are then linearized for small anisotropies.
For completeness we recall the Lema{\^i}tre-Hubble diagram for small anisotropies \cite{vts}.

Section \ref{dyn} contains the linearized Einstein equations in small anisotropy and small curvature and their solutions. Then we use the exact solutions of Friedman's equation given in the appendix to linearize the Lema{\^i}tre-Hubble  diagram and the drift simultaneously in small anisotropies and small curvatures.  The domain of validity of linearizing anisotropies must be controlled carefully because they tend to infinity with growing redshift by Wald's theorem \cite{wald}.

The appendix recalls the solutions of the Friedman equation with positive cosmological constant and comoving dust in terms of Jacobi elliptic functions. We linearize these solutions for small curvature.  The results of the appendix are used in section \ref{dyn}.

 \section{Kinematics in
Axial Bianchi IX universes \label{kin}} 

Axial Bianchi IX universes have two scale factors $a(t)$ and $c(t)$:
 \begin{align}
\de \tau^2= \de t^2&-{\textstyle\frac{1}{4}} a^2 \de x^2-{\textstyle\frac{1}{4}}(a^2\cos^2x+c^2 \sin^2x)\,\de y^2
-{\textstyle\frac{1}{4}}c^2\,\de z^2-{\textstyle\frac{1}{2}} c^2\sin x\,\de y\,\de z\,.
\label{metricaac}
\end{align}
Thanks to Emmy Noether's theorem the four Killing vectors of axial Bianchi IX universes produce four conserved quantities on all geodesics $x^\mu (q)$. Denoting by an over-dot the derivative with respect to the affine parameter $q$ we can write these conserved quantities \cite{vts}:
\bb
\aaa&\dpp=&a^2
(
 \sin y\,\dot x
 -{\sin x}\,{\cos x}\,\cos y\,\dot y)
 +c^2\,{\cos x}\,\cos y\,
 (\sin x\, \dot y+\dot z)\,,\label{conserved1}\\
\aaab &\dpp=& a^2(
 \cos y\,\dot x
 +{\sin x}\,{\cos x}\,\sin y\,\dot y)
 -c^2\,{\cos x}\,\sin y\,
 (\sin x\, \dot y+\dot z)\, ,\\
\bbb&\dpp=&\hspace{2.7cm}a^2\cos^2x\,\dot y
+c^2\sin x\,(\sin x\,\dot y+\dot z),
\label{conserved3}\\
\ccc&\dpp=& \hspace{5.2cm}c^2(\sin x \,\dot y+\dot z)\,,\label{conserved4}
\ee
 or equivalently,
\bb
\dot x &=&\pm\,\frac{1 }{a^2}\lb
\aaa^2+\aaab^2-\lp \frac{\bbb\sin x-\ccc}{\cos x} \rp^2\rb^{1/2}\,,
\label{xdot}\\[2mm]
\dot y&=&\,\frac{\bbb-\ccc\sin x}{a^2\cos^2 x}\, ,\label{ydot}\\[2mm]
\dot z&=&\,\frac{\ccc}{c^2}\, -\sin x\,\frac{\bbb-\ccc\sin x}{a^2\cos^2 x}\, ,
\label{zdot}\\[2mm]
 \aaa\cos y-\aaab \sin y &=& -\,\frac{\bbb\,\sin x-\ccc}{\cos x}\, \label{star}. 
\ee
For light-like geodesics we have in addition $\dot x^\mu g_{\mu \nu }\dot x^\nu=0$,  or:
\bb
  \dot t^2\,\ =\ \,\frac{\aaa^2+\aaab^2+\bbb^2-\ccc^2}{4\,a^2}\,+\,\frac{\ccc^2}{4\,c^2}\,\   =\,\
  \,\frac{a^2}{4}\,\lp\dot x^2+\cos^2 x\,\dot y^2\rp
  +\,\frac{c^2}{4}\, \lp \sin x\, \dot y+\dot z\rp^2
  \label{tdot} .
 \ee
 
 \subsection{
Drift of comoving extra-galactic sources}

 At every comoving point (of observation) in any axial Bianchi I universe there is a privileged direction, constant in time, in which there is no drift, and all directions in the plane orthogonal to this privileged direction are also driftless. Let us show that this property also holds in axial Bianchi IX universes.
 
 Since all Bianchi universes are homogeneous in the sense that any two points are related
by an isometry, it is sufficient to show the above property in a particular point. We choose
the position of our comoving observer at the equator of the 3-sphere, $\vec x_0=0$. Today, $t=t_0$, she observes a comoving light source in a direction ${-\,\dot {\vec x}}(t_0)$.
 
 Let us evaluate the Noether constants there,
 \bb
 \aaa_0=c_0^2\,\dot z_0,\qq\aaab_0=a_0^2\,\dot x_0,
 \qq \bbb_0=a_0^2\,\dot y_0,\qq \ccc_0=c_0^2\,\dot z_0.
 \ee
 We have added the subscript $\cdot_0$ to the Noether constants, because later we will evaluate them at the point of emission.
 
 To prove that the $z$-direction is the privileged direction at this point we set $\dot x_0=\dot y_0=0$ and $\dot z_0\not=0$.  
 Then equation (\ref{xdot}) reads,
 \bb
 \dot x &=&\pm\,\frac{c_0^2 }{a^2}\,\dot z_0\lb
1-\frac{1}{\cos^2 x} \rb^{1/2}.
\ee
Therefore $x(q)=0$ for all times, otherwise $\dot x$ would become imaginary. Then equation (\ref{ydot}) tells us that also $y$ remains zero and  equation (\ref{zdot}) gives 
\bb
\dot z =\,\frac{c_0^2}{c^2}\,\dot z_0\,. 
\ee
Ten years later, $t=t_0+T_D$, she observes the same light source still in the $z$-direction, because 
\bb
\vec x'\dpp=\,\frac{\dot {\vec x}}{\dot t}\, =
\begin{pmatrix}
0\\0\\z'
\end{pmatrix} =
\begin{pmatrix}
0\\0\\2/c
\end{pmatrix} =\dpp
\vec u_z
,
\ee
is a {\it unit} vector with respect to the negative of the spatial part of the metric (\ref{metricaac}). Indeed we have set the speed of light to one. 

Next let us consider the orthogonal plane in the tangent space of our observer at $\vec x_0=0$ where the spatial part of the metric (\ref{metricaac}) is Euclidean. Therefore the orthogonal plane is characterized by $\dot z_0=0$. Then equations (\ref{xdot}-\ref{zdot}) reduce to:
\bb
\dot x&=&\pm\,\frac{a_0^2}{a^2}\,\dot x_0 \lb 1-
\,\frac{\sin^2 x}{\cos^2x}\rb^{1/2},
\\[2mm]
 \dot y&=& \,\frac{a_0^2}{a^2}\,\dot y_0\,,
 \\[1mm]
 \dot z&=& -\,\frac{a_0^2}{a^2}\,\dot y_0\,\frac{\sin x}{\cos^2x}\,.
 \ee
Note that the second scale factor $c(t)$ has disappeared from these geodesic equations. Therefore these geodesics are identical to the ones on the 3-sphere, which are driftless for all values $(\dot x_0,\dot y_0)$.

If the comoving observer is situated anywhere else (except at the North pole, $x=3 \pi /2$, or at the South pole, $x=\pi /2$) a similar, though longer calculation shows that the privileged direction remains the z-direction. We skip this calculation because our results in subsection \ref{subdrift} constitute an alternative proof.

 \subsection{Trajectory of a first photon}
 
 Now we choose the emitter at the North pole, $x_e={\textstyle\frac{3}{2}} \pi $, $z_e=y_e\pm\pi $. This  simplifies the computations, but we must pay attention to the coordinate singularity there. 
 
 Consider a photon sent off at cosmic time $t_e=t(q_e)$ with initial conditions:
\bb
\begin{tabular}{llll}
 $t(q_e)=t_e$,&  $x(q_e)=x_e={\textstyle\frac{3}{2}} \pi$,&
   $y(q_e)=y_e$,&$z(q_e)=z_e=y_e+\pi$,
 \\[2mm]
 $\dot t(q_e)=\dot t_e>0$,& $\dot x(q_e)=\rrr>0$,&
$\dot y(q_e)=\sss$,& 
 $\dot z(q_e)=\dot z_e$.
 \end{tabular}\label{initial}
 \ee
 Then  by equations (\ref{conserved1})-(\ref{conserved4}) we have:
\bb 
\,\frac{\aaa}{\sin y_e}\, =\,\frac{\aaab}{\cos y_e}\,=a_e^2\,\rrr,\qq
\aaa^2+\aaab^2=a_e^4\,\rrr^2,\qq
\bbb=-\ccc=c_e^2(\sss-\dot z_e),
\ee
and from equation (\ref{zdot}),
\bb
\dot z_e=\lb 1-2\,\frac{a_e^2}{c_e^2}\rb \sss.
\ee 
We introduced the abbreviations  $a_e\dpp=a(t_e)$, $c_e\dpp=c(t_e)$.
Other short-hands will be used:  
\bb 
\ssss\dpp=\,\frac{2\sss}{\rrr}\,,\qq\qq&& f(x)\dpp=\,\frac{1+\sin x}{\cos x}\,,\qq
f_x\dpp=\,\frac{\de}{\de x}\,f,\\[2mm]
W(t)\dpp=2\lb \frac{1}{a^2}\,+\,\frac{\ssss^2}{c^2}\rb^{-1/2},
&&
V(x)\dpp = \lb 1-\ssss^2f(x)^2\rb^{-1/2}.\label{4defs}
\ee
Then the geodesic equation yields
 the relation,
\bb
\sin(y-y_e)\ =\ \ssss\,f(x), \label{sstar}
\ee
 and the velocity vector of the photon,
\bb
\vec x'=\,\frac{W}{a^2}\, 
\begin{pmatrix}
V^{-1}\\[2mm]
\ssss\,f_x\\[2mm]
-\ssss\lp f_x+\lb{a^2}/{c^2}-1\rb\rp
\end{pmatrix},\ee
which is a unit vector on the celestial sphere, because we have set the speed of light to unity. 

If the photon arrives today, $t_0=t(q_0)$, in our telescope situated at $\vec x_0\dpp=\vec x(q_0)$, we can integrate once more by separation of variables and obtain,
 \bb
 \int_{t_e}^{t_0}\,\frac{W}{a^2}\, =\int_{x_e}^{x_0}V
 =\,\frac{2}{\sqrt{1+\ssss^2}}\, \arctan(\sqrt{1+\ssss^2}\,f_0\,V_0),\label{firstin}
\\[2mm]
 \,\frac{s}{2}  \int_{t_e}^{t_0}\lb\frac{1}{a^2}\,- \,\frac{1}{c^2}\rb W
 =\,\frac{y_0+z_0}{2}\, -y_e-{\textstyle\frac{1}{2}} \pi .\label{secondin}
 \ee

 \subsection{Trajectory of a second photon and redshift}
 
Have a second photon emitted from the North pole at cosmic time $t_0+T$, where $T$ is an atomic period. Suppose this second photon arrives in our comoving telescope still at $\vec x_0$ at cosmic time $t_0+T_D$. ($\cdot_D$ stands for Doppler.) Assume that the atomic periods, $T$ and $T_D$, are much smaller than the `time of flight' $t_0-t_e$. Also suppose that the geodesics of the two photons, $x^\mu (q)$ and $\tilde x^\mu (q)$, $q\in[q_e,q_0]$ be infinitesimally close, with the second geodesic's initial conditions at $q_e$:
\bb
\begin{tabular}{llll}
 $\tilde t(q_e)=t_e+T$,&  $\tilde x(q_e)=x_e={\textstyle\frac{3}{2}} \pi$,&
   $\tilde y(q_e)=\tilde y_e$,&$\tilde z(q_e)=\tilde y_e+\pi$,
 \\[2mm]
 $\dot {\tilde t}(q_e)>0$,& $\dot {\tilde x}(q_e)=\tilde \rrr>0$,&
$\dot{\tilde y}(q_e)= \ssst$,& 
 $\dot {\tilde z}(q_e)=\lb 1-2\, a^2_{\tilde e}/ c^2_{\tilde e}\rb  \ssst$,
 \end{tabular}
\ee
with $a_{\tilde e}\dpp =a(t_e+T)$.
Then we must adjust the initial direction
 of the second photon,
 \bb
 \tilde y_e=\dpp y_e[1+\delta  ],\qq\qq
  \tilde \ssss\dpp=\,\frac{2 \ssst}{\tilde \rrr}\,= \dpp \ssss[1+\epsilon ],
 \ee
  such that its final comoving position coincides with that of the first photon, $\, \tilde {\vec x}(t_0+T_D)=\dpp\tilde {\vec x}_{\tilde 0}= \vec x_0$. This adjustment is infinitesimal. Therefore we may keep only linear terms in $T,\,T_D,\,\delta  $ and $\epsilon $. In this approximation equation (\ref{sstar}) yields
\bb
\delta \cdot y_e\sim-\epsilon \,\frac{\ssss f_0}{\sqrt{1-\ssss^2f_0^2}}\,=-\epsilon \ssss f_0V_0. 
\ee
We need three more short-hands:
  \bb
 W_\epsilon&\dpp=&2\lb \frac{1}{a^2}\,+\,\frac{\tilde \ssss^2}{c^2}\rb^{-1/2}
 \sim W\lb1-\,\frac{\epsilon \ssss^2}{4c^2}\,W^2\rb,
  \label{Weps}\\[2mm]
  V_\epsilon&\ \dpp  =&\qq  \lb 1-\tilde \ssss^2f^2\rb^{-1/2}\ \sim V\lb1+\epsilon \ssss^2f^2V^2\rb,
  \label{Veps}
  \\[2mm]
   I_3(t)&\dpp=&{\textstyle\frac{1}{4}} \int_{t _e}^t\frac{W^3}{a^2}\lb\frac{1}{a^2}-\,\frac{1}{c^2}\rb.
 \label{I3}
  \ee 
Then the two equations (\ref{firstin}), (\ref{secondin}) yield for the second photon,
\bb
\,\frac{1}{a_0^2}\,W_0\,T_D- \,\frac{1}{a_e^2}\,W_e\,T&\sim&
 \epsilon \ssss^2\lp {\textstyle\frac{1}{4}} \int_{t_e}^{t_0}\,\frac{W^3}{a^2c^2}\, +\int_{x_e}^{x_0}f^2V^3\rp,\label{infx}\\[3mm]
\hspace{-7mm}
\lb\frac{1}{a_0^2}\,-\,\frac{1}{c_0^2}\rb W_0\,T_D- \lb\frac{1}{a_e^2}\,-\,\frac{1}{c_e^2}\rb W_e\,T&\sim&
 \epsilon \la -I_3(t_0) +2  f_0V_0\ra.
 \label{inffi2} 
\ee
Let us take $(1+\ssss^2)$ times equation (\ref{infx}) and subtract $\ssss^2$ times 
equation (\ref{inffi2}). We obtain,
\bb
4\lp\frac{T_D}{W_0}\,- \,\frac{T}{W_e}\rp\sim 0.
\ee
This equation is the direction dependent redshift in axial Bianchi IX universes, 
\bb 
 \zzz\dpp=\,\frac{T_D-T}{T}\,\sim  \,\frac{W_0}{W_e}\,-1,\label{redshift}
 \ee
 of course 
 the redshift $\zzz$ should not be confused with the third Euler angle $z$.

So far we have only used one combination of the two integrals (\ref{infx}) and (\ref{inffi2}) and we have obtained the redshift. In the next subsection we will use the remaining information in the form of equation (\ref{inffi2}) to obtain the drift.

 \subsection{Trajectory of the second photon and drift \label{subdrift}}
 
 Now we interpret $T_D$ not as a redshifted atomic period, but as the time allotted  for observing our source at the North pole. Equation (\ref{inffi2}) will give us the apparent drift during the allotted time $T_D$. 

For the first and second photon we have respectively:
\bb
\vec x'=\,\frac{W}{a^2}\, 
\begin{pmatrix}
V^{-1}\\[2mm]
\ssss\,f_x\\[2mm]
-\ssss\lp f_x+\lb{a^2}/{c^2}-1\rb\rp
\end{pmatrix},\qq
 \tilde {\vec x}'=\,\frac{W_\eps}{a^2}\, 
\begin{pmatrix}
V_\eps^{-1}\\[2mm]
\tilde \ssss\,f_x\\[2mm]
-\tilde \ssss\lp f_x+\lb{a^2}/{c^2}-1\rb\rp
\end{pmatrix},
\ee
 with $\eps$ given by equation (\ref{inffi2}). Both are unit vectors on the celestial sphere. 
 
 Let us compute the drift in {\it four steps:}
 
 Our {\it first step} is to evaluate the arriving direction of the  second photon at its arrival time $t_{\tilde 0}=t_0+T_D$ to first order in $\eps$ and $T_D$:
 \bb
 \tilde {\vec x}'_{\tilde 0}\sim \,\frac{W_0}{a^2_0}\, 
\begin{pmatrix}
\hspace{19mm}
V_0^{-1}\lb1+\eps\,e_x+T_D\,\tau\rb
\\[2mm]
\hspace{19mm}
\ssss\,f_{x0}\lb1+\eps\,S_2+T_D\,\tau\rb
\\[2mm]
- \ssss\lp f_{x0}+\lb{a^2_0}/{c^2_0}-1\rb\rp
\lb1+\eps\,S_2+T_D\,(\tau +\tau_z)\rb
\end{pmatrix},
\ee
with
\bb
e_x\dpp=-\ssss^2\la \frac{W_0^2}{4c_0^2}\,+f_0^2V_0^2\ra\, ,
&&
\tau\dpp=\,\frac{W_0^2}{4}\lp\frac{H_0}{a_0^2}\,+\ssss^2\,\frac{H_{c0}}{c_0^2}\rp-2H_0,
\\[2mm]
S_2\dpp=\,1-\,\frac{\ssss^2}{4}\,\frac{W_0^2}{c_0^2},
\hspace{15mm}
&&
\tau_z\dpp=\,2\,\frac{a_0^2/c_0^2\,(H_0-H_{c0})}{f_{x0}+[a_0^2/c_0^2-1]}\,.     
 \ee
 The computation of these four coefficients is straight-forward, but long.
 The interested reader is invited to calculate one of these coefficients in order to appreciate the usefulness of the many abbreviations introduced and of their intermediary calculations. Subsection A.3 of the appendix is also a source of good exercises (no sweat, no fun).
 
 In our {\it second step} we compute the arriving angle of the first photon. Let $\theta $ be the angle between $  {\vec x}'_{0}$ and the privileged direction $\vec u_z$ and denote by ${\vec x}'_{ 0}\,\cdot\,\vec u_z$ the scalar product in 3-space with respect to the negative of the spatial part of the metric (\ref{metricaac}).
 Then
 \bb
 \cos \theta\, =\,  {\vec x}'_{ 0}\,\cdot\,\vec u_z
 =\,\frac{\ssss}{2}  \,\frac{c_0W_0}{a_0^2}
 \la \sin x_0\,f_{x0} -(f_{x0}+[a_0^2/c_0^2-1]) \ra
 =\,\frac{\ssss}{2}  \,\frac{W_0}{c_0}, 
 \label{costh}
 \ee
 implying $S_2=\sin^2 \theta$.
  
 In our {\it third step} we compute the arriving angle of the second photon. Let $\tilde\theta $ be the angle between $ \tilde {\vec x}'_{\tilde 0}$ and the privileged direction $\vec u_z$. 
Then
 \bb
 \cos \tilde\theta\, =\, \tilde {\vec x}'_{\tilde 0}\,\cdot\,\vec u_z\,
 \sim\,\cos \theta \lb 1+\eps\,\sin^2\theta +T_D
 \la \cos^2\theta  \,\frac{1}{\ssss^2}\,\frac{c_0^2}{a_0^2}\,H_0-\sin^2\theta \,H_{c0} \ra 
 \rb.
\ee

In the {\it fourth step} we use equation (\ref{inffi2}) to eliminate $\eps$ in favour of $T_D$,
\bb
\eps\,\sim\,T_D\,\frac{\lb1/a_0^2 -1/c_0^2\rb W_0-
\lb1/a_e^2 -1/c_e^2\rb W_e/(
 \zzz+1) 
}{-I_{30}+2f_0V_0}\,. 
\label{epsTD}
\ee

\subsection{Drift with small anisotropy (still without Einstein's equations) \label{subdriftinf}}

Now let us assume a small anisotropy $\eta(t)$ defined by
\bb
a=a_F\,[1-{\textstyle\frac{1}{2}}  \eta],\qq\qq
c= a_F\,[1+ \eta],
\ee 
or equivalently,
\bb
a_{F}={\textstyle\frac{2}{3}}a+{\textstyle\frac{1}{3}}c,
\qq\qq
\eta=\,\frac{c-a}{a+{\textstyle\frac{1}{2}} c}\,.
\ee
For $\eta=0$ we have $c=a=a_F$ and we are back in a  {\it Friedman} universe with scale factor $a_F$ and positive curvature.

In order to linearize the drift in $\eta$ we need a few intermediary results and start by linearizing $W(t)$ defined by the second of equations (\ref{4defs}),
\bb
W\sim\,\frac{2}{\sqrt{1+\ssss^2}}\,a_F\lb1-{\textstyle\frac{1}{2}} \,\frac{1-2\ssss^2}{1+\ssss^2}\,\eta\rb+O(\eta^2).  \label{linW}
\ee
From this we obtain the linearized redshift,
\bb
 \zzz+1&\sim&\frac{a_{F0}}{a_{Fe}}\,\lb 1- {\textstyle\frac{1}{2}} \,\frac{1-2\ssss^2}{1+\ssss^2}\,(\eta_0-\eta_e)\rb,
\ee 
as well as
\bb
\lb\frac{1}{a^2}\,-\,\frac{1}{c^2}\rb&\sim&
\frac{3\eta}{a_F^2}\, ,
\\[2mm]
  \lb\frac{1}{a_0^2}\,-\,\frac{1}{c_0^2}\rb W_0 -
   \lb\frac{1}{a_e^2}\,-\,\frac{1}{c_e^2}\rb \frac{W_e}{ \zzz+1}\, &\sim&
\frac{6}{\sqrt{1+\ssss^2}}\, \frac{\eta_0-\eta_e}{a_{F0}}\, ,
\ee 
and the integral $I_3(t)$, equation (\ref{I3}),
\bb
I_3\sim \,\frac{6}{(1+\ssss^2)^{3/2}}\,\int_{t_e}^t\,\frac{\eta}{a_F}\,.
 \ee
 
 Now let us linearize equation (\ref{epsTD}) in $\eta$. Here we need $f_0V_0$ to zeroth order in $\eta$, i.e. for $a=c=a_F$, the Friedman case. Evaluating equation (\ref{firstin}) in this case, we obtain
 \bb
 f_0V_0= \,\frac{1}{\sqrt{1-\ssss^2}}\, \tan \chi _{e0}\,+O(\eta),\qq\qq
 \chi _{e0}\dpp =\int_{t_e}^t\,\frac{1}{a_F}\,,\ee
and then,
\bb
\eps\,\sim\,3\,\frac{\eta_0-\eta_e}{a_{F0}\tan \chi _{e0}}\,T_D.
\ee 
We also need $\cos\theta $ in the Friedman case and equation (\ref{costh}) yields,
\bb
\cos\theta _F\,=\,\frac{-\ssss}{\sqrt{1+\ssss^2}}\, .
\ee
Finally we can write the drift,
\bb
\delta \,\cos \theta\,\dpp =\,
\cos \tilde \theta -\cos\theta\,=\,
3 \sin^2\theta \,\cos\theta 
\la \,\frac{\eta_0-\eta_e}{a_{F0}\tan \chi _{e0}}\,-{\textstyle\frac{1}{2}} \eta'_0\ra T_D\,+\,O(\eta^2).
\ee 

\subsection{
The Lema{\^i}tre-Hubble diagram with small anisotropy
\\  ${}$\hspace{35mm} (without Einstein's equations)}

For comparison we recall \cite{vts} the formulae of the Lema{\^i}tre-Hubble diagram, i.e. the apparent luminosity $\ell$  versus redshift as a parametric plot obtained by eliminating the emission time $t_e$, in axial Bianchi IX universes in the linear approximation of small anisotropy $\eta(t)$:
\bb
\zzz+1
&\sim&\,\frac{a_{F0}}{a_{Fe}}\,\lb 1- \,\frac{1-3\cos^2\theta}{2}\, (\eta_0-\eta_e)\rb, 
\label{redshift2}\\[2mm]
\ell&\sim& \ell_F
\lb 1+\,\frac{1-3\cos^2\theta}{2}\,\lp
 \eta_0-5\eta_e+4\,\frac{\chi _{e0}}{\tan\chi_{e0} }\,\bar\eta_{e0} \rp    \rb, 
 \label{apLum}
\ee
with
\bb
\ell_F=\,\frac{L}{4\pi\,a_{F0}^2\,\sin^2\chi_{e0} }\lp\frac{a_{Fe}}{a_{F0}}\rp^2,
\ee
with the absolute luminosity $L$ of the supernova and with the definition:
\bb
\bar\eta_{e0}\dpp=\,\frac{1}{\chi_{e0}}\,  \int_{t_e}^{t_0}\frac{\eta}{a_F}\, \,.
\ee

 \section{Dynamics in
Axial Bianchi IX universes \label{dyn}}  

Axial Bianchi IX universes have two scale factors, $a(t)$ and $c(t)$, 
  two Hubble parameters,
\bb
H\dpp =\,\frac{a'}{a}\,, \qq\qq H_c\dpp =\,\frac{c'}{c}\,.
\ee
For small anisotropy $\eta(t)$ we obtain to first order,
\bb 
H\sim H_F-{\textstyle\frac{1}{2}} \eta',
\qq
H_c\sim H_F+\eta',
\qq
H_{F}\sim {\textstyle\frac{2}{3}} H+{\textstyle\frac{1}{3}} H_{c},
\qq\qq
\eta'\sim \textstyle\frac{2}{3} (H_{c}-H).
\label{correct}
\ee
(There is a typo in our paper \cite{vts}, the $2/3$ in front of $(H_{c}-H)$ in the last equation here is correct.)

Let us now introduce dynamics in the form of Einstein's equations with positive cosmological constant $\Lambda $ and comoving dust of density $\rho _0$ today.
Then we have
 two Einstein equations, which we write in the following form \cite{vts},
\bb
 3H^2+2H(H_c-H)+\,\frac{1}{a^2}\lp 4-\,\frac{c^2}{a^2}\rp
&=&\Lambda +8\pi\,G\,\rho _0\,\frac{a_0^2c_0}{a^2c},\label{tt9}
\\[2mm]
(H_c-H)'+(2H+H_c)(H_c-H)-
\,\frac{4}{a^2}\lp 1-\,\frac{c^2}{a^2}\rp
&=&0 .\label{xx-zz}
 \ee
 
 Our strategy is to first linearize these Einstein equations in the small anisotropy $\eta(t)$: 
 
 {\bf To zeroth order} in $\eta$ we return to Friedman universes with positive curvature and there we use the explicit solutions summarized in the appendix and take their limit of small (constant, dimensionless) curvature $\nu\dpp=4/(a_{F0}^2\Lambda )\rightarrow 0$.
 
{\bf To first order} in $\eta$ we subtract the Friedman equation from the first Einstein equation (\ref{tt9}) and evaluate the difference today. We find $\rho _0\sim\rho _{F0}$. Linearizing the second Einstein equation (\ref{xx-zz}) yields,
\bb
\eta''+3\,H_F\,\eta'+\,\frac{8}{a_F^2}\,\eta \,=\, 0\,+O(\eta^2). \label{eta9}
\ee
Now for small curvature we also want to linearize this equation in $\nu$. To this end we rewrite the coefficient of the third term as:
\bb
\,\frac{8}{a_F^2(t)}\,=\,\frac{2\Lambda }{s^2(\chi (t))}\, \nu\,=
\,\frac{2\Lambda }{s_f^2(\chi (t))}\, \nu\,+\,O(\nu^2).
\ee
The {\it functions} $s(\chi )$ and $s_f(\chi )$ are the scale factors in curved and flat Friedman universes normalized to be unity today. 
They are defined in the appendix and must not be confused with the initial condition $\ssss\dpp=2\sss/\rrr$.

Then we obtain
\bb
\eta''+3\,H_{Ff}\,\eta' \,=\, 0\,+\, O(\nu^2,\nu\eta,\eta^2), \label{eta9pert}
\ee
with its unique solution
\bb
\eta(t) &\sim&
\eta_0-\,{\textstyle\frac{2}{3}} \,\frac{\eta'_0}{H_{F0}} \,
\frac{\sqrt{\Omega _{\Lambda }}}{1-\Omega _{\Lambda }} 
\lb \coth\lp{\textstyle\frac{1}{2}}  \sqrt{3 \Lambda}\, t\rp-
1/\sqrt{\Omega _{\Lambda }}\,\,
\rb,
\label{etat}
\ee
and in particular
\bb
\eta_0-\eta_e &\sim&
\,{\textstyle\frac{2}{3}} \,\frac{\eta'_0}{H_{F0}} \,
\frac{\sqrt{\Omega _{\Lambda }}}{1-\Omega _{\Lambda }} 
\lb \coth\lp{\textstyle\frac{1}{2}}  \sqrt{3 \Lambda}\, t_e\rp-
1/\sqrt{\Omega _{\Lambda }}\,\,
\rb.
\ee

\subsection{
The Lema{\^i}tre-Hubble diagram in Friedman universes}  

In a spherical Friedman universe the redshift $\zzz$ 
 of a photon emitted at emission time $\chi _e$ and arriving today $\chi _0$ in our telescope is given by,
\bb 
\,\frac{1}{\zzz+1}\,\dpp=\,\frac{a_F(t_e)}{a_{F0}}\,  =\dpp\,
u\,=\,s(\chi _e),\qq \text{or} \qq
\chi _e=\text{Inv}s\lp\frac{1}{\zzz+1}\rp. 
\ee
The emission time in the flat Friedman universe is,
\bb 
\chi _{ef}=\text{Inv}s_f\lp\frac{1}{\zzz+1}\rp. 
\ee
Let us denote by $L$ the absolute luminosity of the emitter. Then the apparent luminosity in a spherical Friedman universe is,
\bb
\ell_F=\,\frac{L}{4\pi\,a_{F0}^2\,\sin^2\chi_{e0} }\lp\frac{a_F(t_e)}{a_{F0}}\rp^2
=\,\frac{L}{4\pi\,a_{F0}^2\,\sin^2\chi_{e0} }\lp\frac{1}{\zzz+1}\rp^2,
\ee
with the conformal time of flight of the photon going between emitter and observer,
\bb
\chi_{e0}\dpp =\int_{t_e}^{t_0} \,\frac{\de t}{a_F(t)}\,=\chi _0-\chi_e=\text{Inv}s(1)-\text{Inv}s\lp\frac{1}{\zzz+1}\rp < \pi .
\ee
On the other hand the apparent luminosity in the flat Friedman universe is,
\bb
\ell_{Ff}\,
=\,\frac{L}{4\pi\,a_{F0}^2\,\chi_{e0f}^2 }\lp\frac{1}{\zzz+1}\rp^2,
\qq\chi_{e0f}\dpp =\chi _{0f}-\chi_{ef} =\text{Inv}s_f(1)-\text{Inv}s_f\lp\frac{1}{\zzz+1}\rp.
\ee

The Hubble parameter $H_F(t)$ and the redshift $\zzz$ are invariant under rescaling of the scale factor $a_F(t)$. This is not true for the dimensionless conformal time $\chi $, but it is true for the dimensionful product $a_{F0}\,\chi $. We may consider the flat limit of the Lema{\^i}tre-Hubble diagram as a rescaling of $a_{F0}=2/\sqrt{\Lambda }/\sqrt{\nu}$ with $\nu\rightarrow 0$ and $(a_{F0}\,\chi_{e0})$ remaining finite. Then
\bb
a_{F0}\,\sin\chi_{e0}&=&\frac{2}{\sqrt{\Lambda} \sqrt{\nu}}\,
\sin\lp  (a_{F0}\,\chi_{e0})\sqrt{\Lambda} \sqrt{\nu}/2 \rp
\\[2mm]
&\sim &\frac{2}{\sqrt{\Lambda} \sqrt{\nu}}\,
\la (a_{F0}\,\chi_{e0})\sqrt{\Lambda} \sqrt{\nu}/2 -{\textstyle\frac{1}{48}}
(a_{F0}\,\chi_{e0})^3\lp \sqrt{\Lambda} \sqrt{\nu}\rp^3
 \ra
 \\[2mm]
 &=&a_{F0}\,\chi_{e0}\lb1-{\textstyle\frac{1}{24}}
(a_{F0}\,\chi_{e0})^2 \Lambda \nu\rb
\\[2mm]
&\sim& a_{F0}\,\chi _{e0f}\lb
1-\la \alpha (\zzz)+ \,\frac{1}{24}\, 
(a_{F0}\,\chi_{e0f})^2 \Lambda\ra
 \nu
\rb,
\ee
  with
\bb
\chi _{e0}\sim\chi _{e0f}\lb 1-\alpha (\zzz)\,\nu \rb,\qq\qq
\alpha (\zzz)\dpp=\,\frac{\text{Inv}s_f(1)\,\beta (1)-\text{Inv}s_f(1/(\zzz+1))\,\beta (1/(\zzz+1))}{\text{Inv}s_f(1)-\text{Inv}s_f(1/(\zzz+1))}\,,
\label{nu_u}\ee
and the function $\beta  (u)$ is given in the appendix, equation (\ref{betay}).
By l'H\^opital's rule $\alpha (\zzz)$ is regular today, $\zzz=0$. For example with $\Omega _\Lambda =0.7$ and $-\Omega _\kappa =0.05$, we have 
$\nu=9.5\,\%$,
 $\alpha (0)=0.582$ 
 and then the function $\alpha (\zzz)$  varies monotonically and saturates rapidly at $0.803$ with growing redshift. 

Finally,  to first order in $\nu$, the Lema{\^i}tre-Hubble diagram in Friedman universes as a function of redshift $\zzz$  reads,
\bb
\ell_{F}\,
\sim\,\ell_{Ff}
\lb
1+2\la \alpha (\zzz)+ \,\frac{1}{24}\, 
(a_{F0}\,\chi_{e0f})^2 \Lambda\ra \nu
\rb.
\ee

\subsection{Lema{\^i}tre-Hubble diagram and drift in axial Bianchi IX, \\${}$ \hspace{22mm}
 small anisotropy and small curvature  }

Now we are ready to also linearize equation (\ref{apLum}) in the limit of small curvature,
\bb
\ell&=&\frac{L}{4\pi\,a_{F0}^2\,\chi_{e0f}^2 }\lp\frac{1}{\zzz+1}\rp^2\,
\bigg[ 1
+2\la \alpha (\zzz)+ \,\frac{1}{24}\, 
(a_{F0}\,\chi_{e0f})^2 \Lambda\ra \nu
\\&&
\hspace{44mm}
+\,\frac{1-3\cos^2\theta}{2}\,\lp
 \eta_0-5\eta_e+4\,\bar\eta_{e0} \rp    \bigg]+\,O(\nu^2,\nu\eta,\eta^2). 
 \label{ellB9}\nonumber
\ee
Note that the small curvature $\nu$ and the small anisotropy $\eta$ appear `additively'. A numerical indication of this `additivity' can be found in figure 2 of our reference \cite{vts}. (Note another typo in the explanation of this figure: the values of $\Omega _{\Lambda 0}$ and $\Omega _{m 0}$ have been mixed up.)
 Setting $\nu=0$ we recover the result from axial Bianchi I \cite{stv}.

For the drift the situation is different,
\bb
\delta \,\cos \theta&=&
\frac{3}{2}\,  \sin^2\theta \,\cos\theta \, \eta'_0
\la 
\frac{4}{3H_{F0}} 
\,
\frac{\sqrt{\Omega _{\Lambda }}}{1-\Omega _{\Lambda }}
\lb
 \coth\lp{\textstyle\frac{1}{2}}  \sqrt{3 \Lambda}\, t_e\rp-
1/\sqrt{\Omega _{\Lambda }}\,\,
\rb-a_{F0} \chi _{fe0} 
\ra \frac{T_D}{a_{F0} \chi _{fe0}}
\nonumber\\[2mm]
&& \hspace{33mm}
+\,O(\nu^2,\eta^2,\nu\eta). 
\label{delcos}
\ee 
Up to second order terms this drift is equal to the one in axial Bianchi I \cite{Mar,drift1}.

In linear approximation we may add the effect of a small peculiar velocity $\vec V$ of the observer in a flat Friedman univers  \cite{drift1,makarov} (we have set the speed of light to one),
\bb
\delta \,\cos \theta&=&
\frac{3}{2}\,  \sin^2\theta \,\cos\theta \, \eta'_0
\la 
\frac{4}{3H_{F0}} 
\,
\frac{\sqrt{\Omega _{\Lambda }}}{1-\Omega _{\Lambda }}
\lb
 \coth\lp{\textstyle\frac{1}{2}}  \sqrt{3 \Lambda}\, t_e\rp-
1/\sqrt{\Omega _{\Lambda }}\,\,
\rb-a_{F0} \chi _{fe0} 
\ra \frac{T_D}{a_{F0} \chi _{fe0}}
\nonumber\\[2mm]
&& \hspace{29mm}
-\sin\theta\, \bigg\{\cos\theta \,(\cos\varphi  \,V_x+\sin\varphi  \,V_y)+\sin\theta \,V_z \bigg\}
\,\frac{T_D}{a_{F0} \chi _{fe0}}
\nonumber\\[2mm]
&& \hspace{33mm}
+\,O(\nu^2,\eta^2,V^2,\nu\eta,\nu V,\eta V), 
\ee 
and 
\bb
\sin \theta \,\delta \varphi =(-\sin\varphi \,V_x+\cos\varphi \,V_y)\,\frac{T_D}{a_{F0} \chi _{fe0}}\,+\,O(\nu^2,\eta^2,V^2,\nu\eta,\nu V,\eta V).
\ee
This addition is the well-known kinematic dipole, which breaks the symmetry generated by the fourth or axial Killing vector $\pa_\varphi $.

We have supposed that all functions encountered are analytic and acknowledge that the domains of convergence must be controlled after the fit by reducing the domain of input parameters as necessary. We will suppose that all our small parameters today, the `Hubble stretch' $\eta'_0/H_{F0}$, the curvature parameter $\nu$ and the  observer's peculiar velocity $\vec V$ (we have set the speed of light to unity)  are small, at most of the order of 10 \%. Then neglecting second-order terms induces errors of the order of 1 \%.

In our past, $t<t_0$ (positive redshift) the anisotropy $\eta(t)$ and the curvature $\nu\alpha (\zzz)$ that come from Einstein's equations must remain perturbatively small. We checked this for the curvature term in the Lema{\^i}tre-Hubble diagram  just after equation (\ref{nu_u}). For the anisotropy, equation (\ref{etat}), Wald's theorem \cite{wald} sets  constraints on the Hubble stretch for any given redshift interval. In our recent work \cite{makarov} we find -- for a fit of axial Bianchi I universes ($\eta_0=0$) to the drift of Gaia quasars with redshift $\zzz<3$ -- a Hubble stretch of $(5.0 \pm 0.7)\,\%$. For $\zzz\in [0,3]$ the anisotropy function $\eta(t)$ varies from 0 to -23 \%, slightly outside the perturbative domain as the redshift approaches 3.  In his fit to the Cosmic Microwave Background ($z\sim 1100$ or $t_e\sim450\,000$ yrs)  Cea \cite{cea1}  finds a Hubble stretch of $(3.8 \pm 1.2) \cdot 10^{-8}$ while $\eta(t)$ varies from 0 to $-0.17\,\%$. 

\section{Conclusions}

Intuitively the equality of the drifts in axial Bianchi I and IX to linear order comes from the fact that for vanishing anisotropy, $\eta=0$, the drifts in both families of universes coincide. This is not true for the Lema{\^i}tre-Hubble diagrams.

Let me say this differently: Suppose that on the right-hand side of the drift equation (\ref{delcos}) there were a non-vanishing term $\nu\,N\,T_D$ in addition to the term in $\eta'_0$. Then in the limit of vanishing Hubble stretch we would obtain a non-vanishing drift in Friedman universes with non-vanishing curvature.

Formally we have proven that for both, drifts and Lema{\^i}tre-Hubble diagrams, the two limits, of small anisotropy and of small curvature, are continuous and commute. (Note that the limit of small curvature is only defined locally for sufficiently short geodesics.) 
This result is non-trivial. Indeed discontinous limits do exist in general relativity, e.g.
the van Dam-Veltman-Zakharov discontinuity arising in the massless limit of massive gravity in certain spacetimes \cite{van,zak}.

On the observational side, drifts cannot distinguish axial Bianchi I from IX universes in linear approximation.
I have not computed second order terms. Nevertheless I doubt that they can cure the cited discrepancy between the CMB- and Gaia-fits.
Therefore we must wait for more and better Gaia results 
\cite{makarov2} 
and supernova 1a observations from the Vera Rubin and James Webb Space Telescopes. 
The latter telescope should be able to observe some super novae with redshift up to 6. 

Meanwhile a fit of the latest Planck data with axial Bianchi IX universes would be welcome.

\vspace{3mm}
\noindent
{\bf Acknowledgements:} 
It is a pleasure to thank Galliano Valent, my sure guide on the Berger sphere.\\

\vspace{2mm}
\noindent
{\bf Data availability:} The data underlying this article will be shared on reasonable request to the corresponding author.\\

\appendix
 
 \section{Exact solutions of Friedman's equation with
\\ ${}$ \hspace{6mm}
 comoving dust in the limit of small curvature
 \label{exact}}

We use notations and results of reference \cite{vts},
\bb
 \Omega _{m}\dpp=\,\frac{8\pi\,G\,\rho _{F0}}{3H_{F0}^2}\,,\qq 
  \Omega _{\Lambda }\dpp=\,\frac{\Lambda }{3H_{F0}^2}\,>0,\qq 
   \Omega _{\kappa }\dpp=\,\frac{-\kappa }{a_{F0}^2H_{F0}^2}\,=
   1-\Omega _{m}-\Omega _{\Lambda }\le 0,  
 \ee
 with $
   \kappa =0\ \text{or}\ 1$,
 \bb
\gamma \dpp=4\,\frac{\Omega _{m }}{\Omega _{\Lambda }}\, ,\qq
\nu\dpp=-\,\frac{4}{3}\, \frac{\Omega _{\kappa  }}{\Omega _{\Lambda }}\, .
\ee 
 We assume that $\gamma ^2-\nu^3$ is positive and  define five more constants,
 \bb 
\tilde s\dpp= -{\textstyle\frac{1}{2}} \lp\gamma +\sqrt{\gamma ^2-\nu^3}\rp^{1/3}
-{\textstyle\frac{1}{2}} \lp\gamma -\sqrt{\gamma ^2-\nu^3}\rp^{1/3},
\\[2mm]
A\dpp=\sqrt{\tilde s^2-\,\frac{3}{4}\, \nu}\qq<\qq
B\dpp=\sqrt{3\tilde s^2-\,\frac{3}{4}\, \nu},\qq\qq 
\sigma \dpp=a_{F0}\sqrt{\frac{\Lambda\,AB}{3}},
\\[2mm]
k^2\dpp=\,\frac{1}{2}\, +\,\frac{3}{4}\,\frac{\tilde s^2-\nu/2}{AB}\,<1.
\hspace{19mm}
\ee
The numbers $A$ and $B$  have nothing to do with the Noether constants $\aaa$ and $\bbb$ in section \ref{kin}.

\subsection{The scale factor as function of conformal time}

We change the independent variable  from cosmic time $t$ to conformal time $\chi $, 
\bb
\chi (t)\dpp=\int_0^t\frac{1}{a_F},\qq 0\leq t.
\label{defchi}
\ee 
The big bang singularity at $t=0$, $a_F(0)=0$, is integrable, $\chi =0$. 

We redefine Friedman's scale factor $a_F(t)$ to be a function of conformal time and normalize it to be unity today,
\bb
s(\chi)\dpp= \,\frac{1}{a_{F0}}\,a_F(t(\chi )), 
\ee
where by abuse of notations we write the inverse function of $\chi (t)$ as 
$t(\chi )$. 

Following Edwards \cite{edw} we can now write the exact solution of Friedman's equation with positive cosmological constant, positive curvature and comoving dust using Jacobi's elliptic function cn,
\bb
s(\chi)\dpp= \,\frac{1}{a_{F0}}\,a_F(t(\chi ))=
|\tilde s|\,\frac{A}{A+B}\,\frac{1-\cn(\sigma \chi,k^2 )}{\cn(\sigma \chi,k^2 )-\cn(\sigma \chic,k^2 )}\,.
 \label{rr1}
\ee
The big bang takes place at $\chi =0$, $t=0$, because cn$\,(0,k^2)=1$.
Note that at finite conformal time $\chic$ the scale factor tends to infinity. The corresponding cosmic time $t(\chic)$ is infinite, indeed the integral
\bb
 \int_0^\infty\,\frac{\de t}{a_F(t)}\, =\dpp\,\chic
\ee
is convergent and
\bb
\chic=\,\frac{1}{\sigma}\,\text{arc}\,\cn \lp\frac{A-B}{A+B},k^2\rp\,
\label{chic}
\ee
is the maximal age of the universe in conformal time.

Let us denote by $\chi =\text{Inv}s(u)$ the inverse function of $u=s(\chi )$, $\chi \in (0,\chic)$, $u\in (0,\infty)$,
\bb
\text{Inv}s(u)\,=\,\frac{1}{\sigma }\,\,\,  \text{arc}\,\cn\lp\frac{(A-B)\,u+|\tilde s|\,A}{(A+B)\,u+|\tilde s|\,A}\,,k^2\rp. \label{invs}
\ee
The age of the universe today in conformal time is given by $\text{Inv}s(1)$:
\bb\label{conf}
\chi _0\,=\,\frac{1}{\sigma }\,\,\,  \text{arc}\,\cn\lp\frac{A\,(1+|\tilde s|)-B}{A\,(1+|\tilde s|)+B}\,,k^2\rp. \label{chi0}
\ee

\subsection{Vanishing curvature}

Let us set $\kappa =0$ and denote flat quantities by the subscript $\cdot _f$,  $\Omega _{\kappa f} =0$. We simplify notations and write $\Lambda _f=\Lambda $, 
$a_{Ff}(t_{0f})=a_{F0}$, $H_{Ff}(t_{0f})=H_{F0}$ and $\Omega _{\Lambda  f} =\Omega _{\Lambda } $.

Using the
 short  hand $R$ for the ratio,
   \bb
R\dpp=\lp\frac{1-\Omega _{\Lambda}}{\Omega _{\Lambda }}\rp^{1/3}=\lp\frac{\Omega _{mf}}{\Omega _{\Lambda }}\rp^{1/3},
 \ee
 (which has nothing to do with the initial condition $\rrr$ of the photon at the North pole, equation (\ref{initial})),
we obtain the following numbers in the flat case:
\bb
k_f^2 &\!\!\!=\!\!\!& 1/2+\sqrt{3}/4=0.933301\cdots,\hspace{4mm}
\\
\gamma _f &\!\!\!=\!\!\!&  4\,R^3,\hspace{16mm}
\nu_f =0,\hspace{16.5mm}
\tilde s_f = -R,\\
A_f &\!\!\!=\!\!\!&  R, \hspace{19.5mm}
B_f = \sqrt{3} R,\hspace{9.5mm}
\sigma_f = 3^{-1/4} R\,a_{F0}\sqrt{\Lambda },
\ee

\subsection{Small curvature and linearization}

Let us introduce a small (positive) curvature, $\nu=4\kappa /a_{F0}^2/\Lambda $ with $\kappa =1$ and large `radius of the universe today' $a_{F0}$. Since in the flat case $a_{F0}$ is a non-essential parameter we wrote $a_{F0}$ for $a_{F0f}$ without confusion now.  We modify only $\Omega _{mf} \rightarrow \Omega _{m}$, but keep  $\Lambda $, $H_{F0}$ and $\Omega _\Lambda $ unchanged. 

We linearize the auxilliary quantities in $\nu$,
\bb
k^2 \!\!\!\!\!&=&\!\!\! k_f^2 +O(\nu^2),\hspace{13mm}
\\[2mm]
\gamma \!\!\!&\sim&\!\!\!  4\,R^3\lb1+\,\frac{3}{4R^3}\, \nu\rb,
\hspace{48mm}
\tilde s \sim -R\,\lb1+\,\frac{1+R}{4R^3}\,\nu \rb,\\[2mm]
A \! \!\!\!&\sim&\!\!\!\! \! R\!\lb1+\frac{-R+2}{8R^3}\, \nu\rb\!\!,\hspace{1mm}
B \sim \!\sqrt{3} R\!\lb1+\frac{R+2}{8R^3}\, \nu\rb\!\!,\hspace{1mm}
\sigma \sim 3^{-1/4} R\,a_{F0}\sqrt{\Lambda }\!
\lb1+\frac{R+2}{8R^3}\, \nu\rb\!\!,
\ee
Note that the difference between $k^2$ and $k_f^2$ is quadratic in $\nu$. Therefore, to linear order we may declare the second argument in Jacobi's elliptic functions to always take the flat value $k_f^2$ and stop writing it explicitly.

Now we are ready to  linearize the inverse  scale factor, equation (\ref{invs}),
\bb
\text{Inv}s(u)\,=\,\frac{1}{\sigma }\,\,\text{arc}\,\cn\,Q(u),\qq \text{with}\qq
Q(u)\dpp=  \,\frac{(A-B)\,u+|\tilde s|\,A}{(A+B)\,u+|\tilde s|\,A}\,, 
\label{invs2}
\ee
 around the inverse scale factor with zero curvature,
\bb 
\text{Inv}s(u ) &\sim& \text{Inv}s_f(u )\,[1+\beta (u)\,\nu].
\label{beta}
\ee
With
\bb
Q_f(u)&\dpp =&\frac{(A_f-B_f)\,u+|\tilde s_f|\,A_f}{(A_f+B_f)\,u+|\tilde s_f|\,A_f}\,=\,\frac{(1-\sqrt{3})\,u+R}{(1+\sqrt{3})\,u+R}\,,
 \\[2mm]
\text{Inv}s_f(u )&\dpp=&
\frac{1}{\sigma_f }\,\,\,  \text{arc}\,\cn\,Q_f(u)\,
=\,\frac{3^{1/4}}{Ra_{F0}\sqrt{\Lambda }}\,\, \text{arc}\,\cn
\,\frac{(1-\sqrt{3})\,u+R}{(1+\sqrt{3})\,u+R}\, 
, 
\label{invsf}
\ee
we obtain,
\bb
\beta (u)\sim\,\frac{1}{8R^2}\,\Big\{&&\!\!\!\!\!\!\!\!\!\frac{R+2}{R}\, 
\\[2mm]
&&-
\frac{Q_f(u)}{\text{arc}\,\cn\,Q_f(u)}\,\frac{((1+\sqrt{3})u+R)^2}{-2u^2+2uR+R^2}
 \,\sqrt{\frac{u}{(u+R)(u^2-Ru+R^2)}}\, 
 (u-1)\,\Big\}
.
\nonumber
\label{betay}
\ee
We are interested in a domain with small positive curvature, $0\le-\Omega_\kappa \le 0.1$. To be specific, let us choose: $\Omega _\Lambda =0.7$, $-\Omega _\kappa =5\,\%$.
Then $\nu=9.5\,\%$
 and the function $\beta (u)$  varies monotonically from $0.582$ to $0.803$ as $u=1/(1+\zzz) $ varies from 0 to 1 (redshift  from $\infty$ to 0).

\end{document}